\RequirePackage{fix-cm}
\documentclass[twocolumn,epjc3]{svjour3}  
\smartqed  
\RequirePackage{graphicx}
\RequirePackage{amsmath}
\usepackage{lineno}
\usepackage{wasysym}

\RequirePackage[colorlinks,citecolor=blue,urlcolor=blue,linkcolor=blue]{hyperref}
\journalname{Eur. Phys. J. A}

\begin{document}
\def\Pb210/{$^{210}$Pb}

\title{Radiopurity of an archeological Roman Lead cryogenic detector}


\author{L.~Pattavina\thanksref{GSSI,TUM,e1,e2}
        \and
        J.W.~Beeman\thanksref{BRK}\and
       M.~Clemenza\thanksref{MIB,UNIMIB}\and
       O.~Cremonesi\thanksref{MIB}\and
       E.~Fiorini\thanksref{MIB,UNIMIB}\and
       L.~Pagnanini\thanksref{MIB,UNIMIB}\and
       S.~Pirro\thanksref{LNGS}\and
       C.~Rusconi\thanksref{LNGS,USC}\and
      K.~Sch\"affner\thanksref{GSSI,LNGS}
}

\thankstext{e1}{also at INFN - Laboratori Nazionali del Gran Sasso}
\thankstext{e2}{Corresponding author: luca.pattavina@lngs.infn.it}

\institute{Gran Sasso Science Institute, I-67100 L'Aquila - Italy\label{GSSI} 
\and
Physik Department, Technische Universit\"{a}t M\"{u}nchen, D-85748 Garching - Germany \label{TUM}
\and
Lawrence Berkeley National Laboratory, CA-94720 Berkeley - USA\label{BRK}
\and
 INFN - Sezione di Milano - Bicocca, I-20126 Milano - Italy\label{MIB}
\and
Dipartimento di Fisica, Universit\`{a} di Milano - Bicocca\label{UNIMIB}
\and
INFN - Laboratori Nazionali del Gran Sasso, I-67100 Assergi (L'Aquila) - Italy\label{LNGS}
\and
Department of Physics and Astronomy, University of South Carolina, SC-29208 Columbia - USA\label{USC}
		}

\date{Received: date / Accepted: date}

\maketitle

\begin{abstract}
Archeological Roman lead (Pb) is known to be a suitable material for shielding experimental apparata in rare event searches. In the past years the intrinsic radiopurity of this material was investigated using different technologies. In this work we applied the latest advancements in cryogenic techniques to study the bulk radiopurity of a 1~cm$^{3}$ sample of archeological Roman Pb. We report the lowest ever measured limit on $^{210}$Pb content in Roman Pb, with a concentration lower than 715~$\mu$Bq/kg. Furthermore, we also studied $^{238}$U and $^{232}$Th impurity concentrations. Our values concur with independent measurements reported in literature. 
\end{abstract}

\section{Introduction}
\label{intro}

The successful realization of ultra-low background experiments searching for elusive natural processes, such as neutrinoless double-beta decay (0$\nu\beta\beta$)~\cite{DBD}, dark matter-nucleus~\cite{DM} interactions or neutrino-nucleus interactions~\cite{CENUS} is particularly difficult due to environmental radioactivity. Unavoidably, natural radioactivity is  included in all materials employed for the detector construction, making it difficult to achieve ultra-low background conditions. As an example, if massive shielding parts surrounding the detector are not made from radiopure materials, they can become themselves the main background source, enlarging the contribution with respect to the external component (i.e. cosmic rays or high energy $\gamma$s). For these reasons material assay is of paramount importance for low counting rate experiments.

The first strategy adopted by ultra-low background experiments is to install their detectors in deep underground laboratories, where they can profit from the overburden rock acting as a 4$\pi$ shielding to suppress the cosmic ray induced component of the background. However, the environmental radioactivity can still limit the detector sensitivity. For this reason, further shields are installed around the experimental set-up and, depending on the size of the detector, different types of materials are employed. Large volume detectors, such as Borexino~\cite{Borex}, Gerda~\cite{Gerda}, DarkSide~\cite{DS}, take advantage of the availability of highly-pure liquids (e.g. water or Ar) for shielding the detectors. At the same time they instrument the shield to function as veto detector. On the other hand, small volume detectors, such as CUORE~\cite{Cuore}, CUPID~\cite{Cupid}, CRESST~\cite{Cresst}, which require a cryogenic facility, are designed to be compact and to be shielded with an overall reduced occupancy, given the technical limitations of the cryogenic systems. It is for this reason that Pb is used as primary shield and it is installed in proximity of the detector's sensitive components.

Pb is also employed as target material for the study of neutrino properties. The two most relevant experiments are: OPERA~\cite{OPERA}, which studied $\nu_{\mu}$ oscillation in matter using Pb and nuclear emulsions as the core part of the detector; and HALO~\cite{HALO} which aims at detecting neutrinos from Supernova explosions in our galaxy using a detector made of Pb and $^{3}$He neutron counters.

The search for highly-pure Pb is, therefore, a topic of interest since many years in the field of low background experiments. While the purification of liquids or gases is a well established technology~\cite{water}, Pb purification is a complex process, due to the natural occurrence of $^{210}$Pb, which is a widely spread radioactive contaminant. 

The most abundant contaminant measurable in Pb samples is $^{210}$Pb, produced by the $^{238}$U decay chain. The reduction of the concentration of $^{210}$Pb in Pb samples is difficult due to their affinity. Whenever a $^{210}$Pb contamination of a Pb sample occurs, most probably this could only be reduced by a cool-down on the time scale of the half-life of the nuclide. Therefore, it can be easily understood that ancient Pb samples could feature an extremely low \Pb210/ concentration, depending on its initial concentration. 

In this work we review the status of different measurements on the radioactivity of ancient Roman Lead samples and we present the results on the excellent radiopurity of a pure Pb cryogenic detector made from archeological Roman Pb.

\section{Archeological Pb}

Archeological Pb is a superior raw material for shielding realization. Even though it was made some time ago its application could still be problematic due to the possible presence of $^{210}$Pb. 

\begin{figure}[t]
\centering
\includegraphics[width=0.5\textwidth]{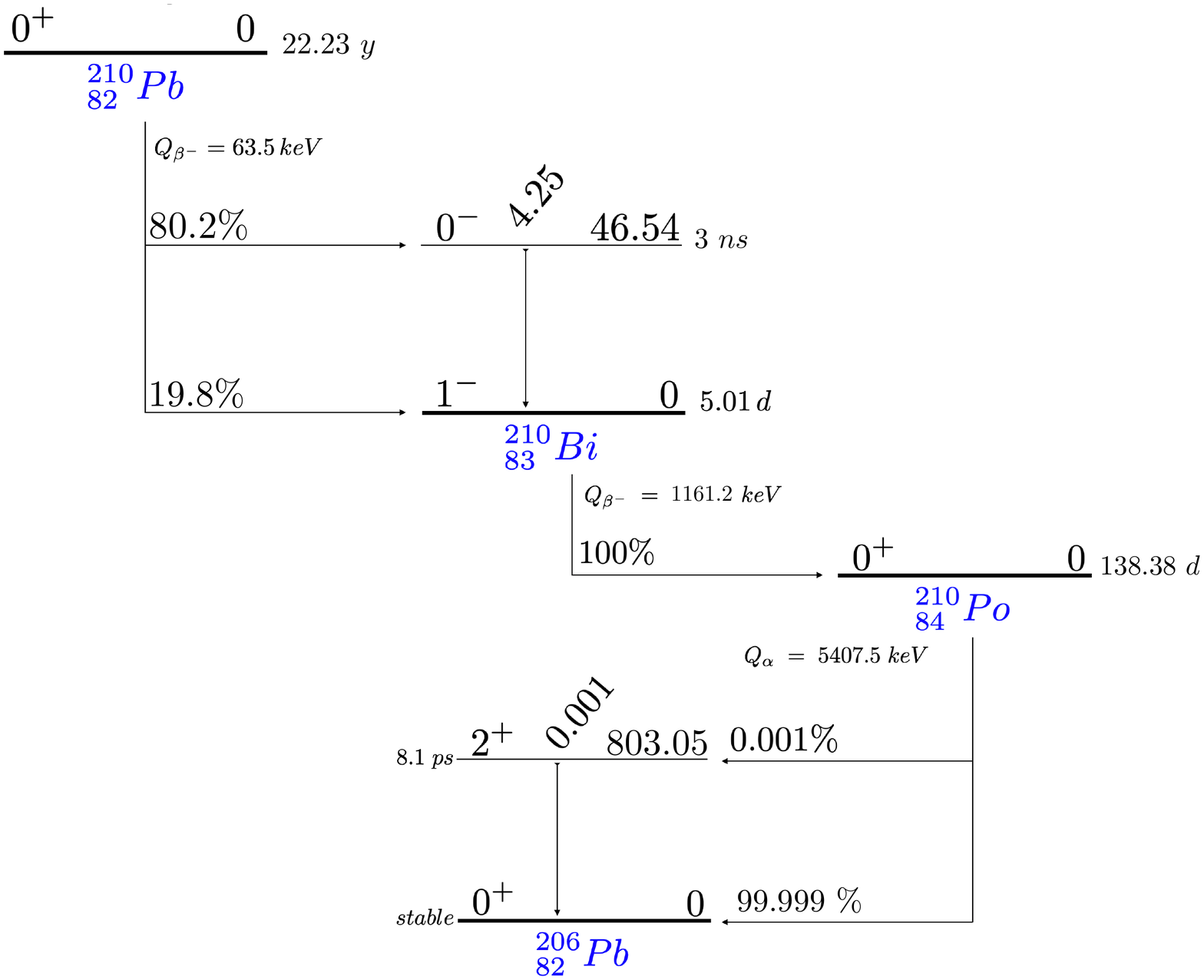}
\caption{Decay scheme of the $^{210}$Pb chain, adapted from~\cite{scheme,Firestone}. For $^{210}$Pb and $^{210}$Bi only the $\beta^-$ branches are shown.}
\label{fig:scheme}
\end{figure}

\Pb210/ can induce different type of signatures in a detector, its decay scheme is shown in Fig.~\ref{fig:scheme}. It decays through a $\beta^-$ channel with an energy transition of 63~keV; it could populate an excited level at 46~keV which would then produce a 46~keV $\gamma$ before reaching the ground state of $^{210}$Bi. This nuclide has a short half-life (5~d), and it decays a through $\beta^-$ channel releasing 1.16~MeV of energy shared among the decay products. The energetic electron can produce bremsstrahlung radiation but also high-energy X-rays (up to 90~keV), while crossing the Pb sample.

After its decay, $^{210}$Bi populates the ground state of $^{210}$Po which, and in most of the cases (99.99\%~\cite{BR}) $\alpha$-decays on the stable nuclide $^{206}$Pb. The energy of the $\alpha$ is about 5.3~MeV, and it is considered a relevant source of background for 0$\nu\beta\beta$ experiments~\cite{Bkg_budget}. In order to mitigate this possible background source, Pb is never directly facing cryogenic detectors but a thick layer of highly-pure Cu is always interleaved between the two.
 

While studying the radiopurity of modern Pb samples and Pb-based compounds, high levels of \Pb210/ concentration can be measured~\cite{Pb210_activity,Pb210_activity2} at the level of tens to hundreds of Bq/kg. This is due to the presence of $^{238}$U decay chain products in Pb ores, as $^{210}$Pb. In fact, when the $^{238}$U decay chain is in secular equilibrium\footnote{All the nuclides of the decay chain have the same activity.}, then $^{210}$Pb is continuously produced from the successive decays of $^{238}$U and its daughter nuclides.

While refining raw metal Pb, a $^{210}$Pb concentration process is encountered, as a matter of fact. In fact, while other radioactive nuclides which are chemically not affine to Pb are segregated from the slag,  $^{210}$Pb is concentrated. This process directly results in high purity Pb, where a high concentration of \Pb210/ could be present. As already discussed, a cool-down time sufficiently long would allow to reduce this residual radioactivity, making Pb an excellent highly-pure material for low background physics. Within the Roman Empire age, Pb was refined for various purposes~\cite{Roman} providing nowadays an important unintentional source of low radioactivity Pb.

The radiopurity of ancient Roman Pb was widely investigated and very low levels of intrinsic contaminations were found. The current best value for $^{210}$Pb is a limit of 4~mBq/kg~\cite{4mBq}, which was obtained using an absorber of pure Pb operated as cryogenic detector. This innovative technique was proposed because of the limited sensitivity of previous measurements. The standard technique used up to the publication time of~\cite{4mBq} was $\gamma$-spectrometry with HP-Ge detectors, which provided limits at the level of hundreds of mBq/kg~\cite{1Bq}. The HP-Ge (High-Purity Germanium) technique is not as sensitive as the cryogenic technique, because of the extremely low detection efficiency due both to geometrical reasons (the detector and the sample are separated) and to self-absorption in the sample. In fact, the most common approach is to detect the 46~keV $\gamma$ which is mostly absorbed by the sample itself. Moreover the emission has a very small branching ratio of 4\% (see Fig.~\ref{fig:scheme}). Instead, the cryogenic technique studies $^{210}$Po, the decay product  $^{210}$Pb. The signal induced by $^{210}$Po is in a favourable region of the energy spectrum with an excellent signal-to-background ratio (at 5.4~MeV), and it also has a branching ratio of almost 100\%. 

Spectrometry with HP-Ge detectors is extremely effective for investigating the concentration of $^{238}$U and $^{232}$Th decay chain products. This is possible thanks to the high energy $\gamma$s produced by these nuclides, (up to 2.6~MeV) which reduce the possibility of self-absorption in the sample, therefore enhancing the detection efficiency. This very same technique, as already discussed, suffers from extremely low detection efficiency for low energy $\gamma$ lines (e.g. $^{210}$Pb $\gamma$ emission).

Currently the best limits on the concentration of $^{238}$U and $^{232}$Th decay chain products are: $<$46~$\mu$Bq/kg and $<$45~$\mu$Bq/kg, respectively~\cite{Radio_rev}. These values are true under the hypothesis that all the elements of the decay chains have the secular equilibrium established in all the parts of the radioactive chain. This is a rather strong assumption given that, depending on how the sample is handled and produced, secular equilibrium could be broken due to the different chemical properties of the decay products.

\begin{table}[]\centering
\begin{tabular}{|c|c|c|c|}
\hline
Detector  & $^{210}$Pb  & Sample mass & Ref.                                                                                                     \\ \hline
          & {[}mBq/kg{]}             & {[}kg{]}    &                                                                                                          \\ \hline
HP-Ge     & $<$900                   & 4.5         & \cite{1Bq} \\ \hline
HP-Ge     & $<$1300                    & 22.1          & \cite{Heusser}                   \\ \hline
Planar Si & 100                      & 0.01        & \cite{Si}             \\ \hline
Bolometer & $<$20                    & 0.09        & \cite{20mBq} \\ \hline
Bolometer & $<$4                     & 0.01        & \cite{4mBq} \\ \hline
\end{tabular}
\caption{Summary of the most sensitive measurements on the concentration of $^{210}$Pb in ancient Pb samples using different techniques.}\label{Tab:tabella}
\end{table}

In Tab.~\ref{Tab:tabella} a summary of the most relevant measurements on low-activity Pb samples is shown.

\section{Archeological Pb based detectors}
Many works can be found in literature where ancient Pb is embedded in cryogenic particle detectors. We can identify two classes of detectors: Pb-based scintillators, such as PbWO$_4$ \cite{PWO} or PbMoO$_4$ \cite{PMO}, and pure-Pb absorbers~\cite{4mBq}.


The possibility to produce detectors, namely scintillating crystals, from radiopure materials is paramount. In fact, a detector with excellent energy resolution and with the ability to perform particle identification can enhance the physics potential of an experiment. Pb-based crystals are known to be excellent scintillators, but unfortunately they suffer from an excess of radioactivity that makes them not suitable for rare event investigations. The reason is the high concentration of $^{210}$Pb which enters in the crystal during its production/growth. For the growth of commercial Pb-based crystals, modern Pb is used, known to feature a high $^{210}$Pb concentration at the level of Bq/kg. Furthermore, such commercial crystals could also exhibit high concentrations of $^{238}$U and $^{232}$Th decay chain products.

Recently, this problem was overcome by the use of ancient Roman Pb for the production of scintillating crystals such as PbWO$_4$ and PbMoO$_4$. Regrettably, these compounds still show a high concentration of $^{210}$Pb, $^{238}$U and $^{232}$Th at the level of hundreds of mBq/kg, due to a not complete control of the crystal growth process.

\begin{figure}[t]
\centering
\includegraphics[width=0.4\textwidth]{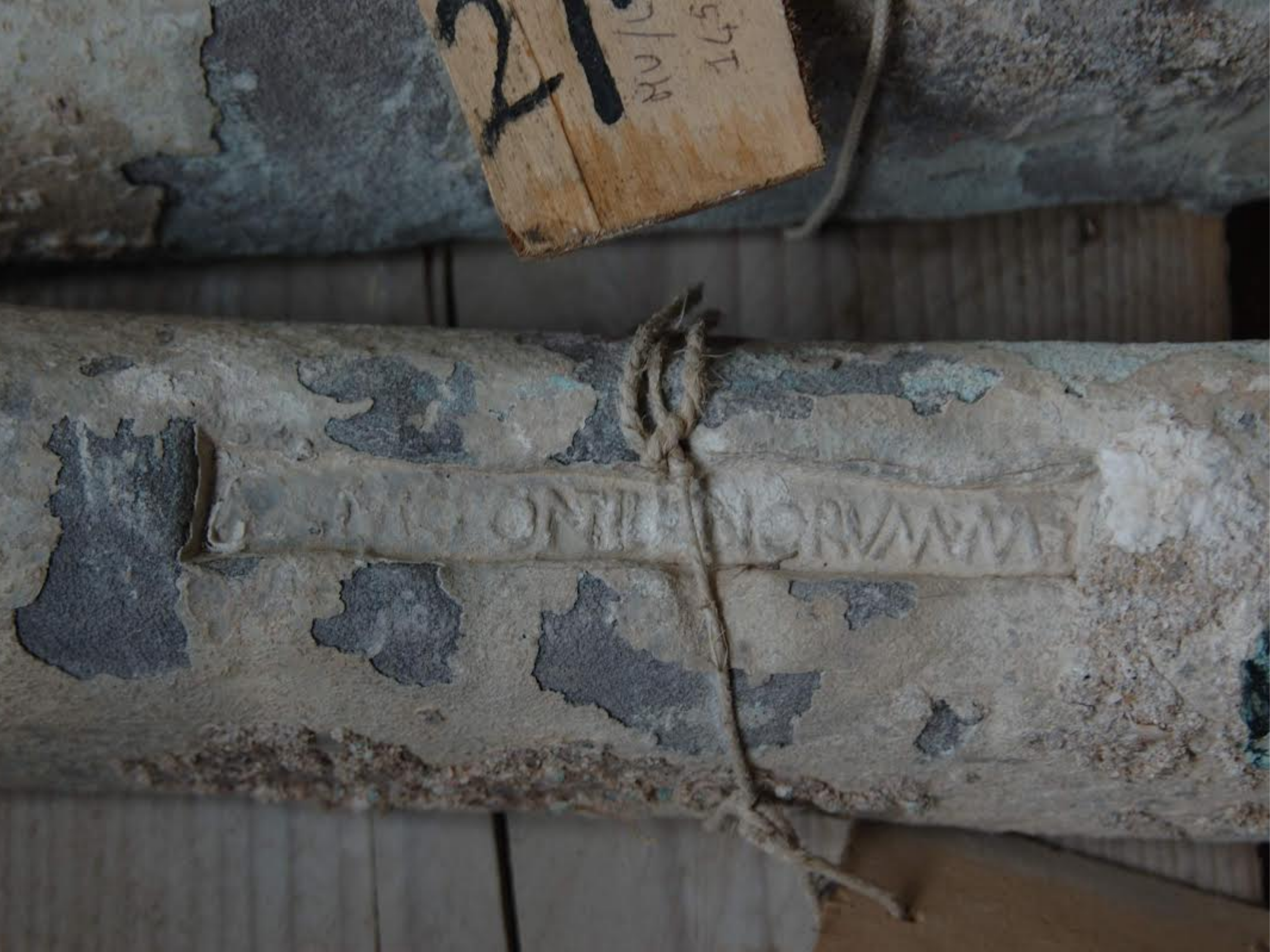}
\caption{Brick of ancient Roman Pb found in the Mediterranean see off the Sardinia shores. The ingot is has a mass of about 23~kg and it is mostly covered by sea salts residuals.}
\label{fig:brick}
\end{figure}

In 1998, pure Pb detectors were also successfully operated~\cite{4mBq} and featured excellent radiopurity levels, thanks to the absence of any production process. The samples were directly cut from the bulk of an ancient Roman Pb brick as the one shown in Fig.~\ref{fig:brick}.

These samples (10~g of mass each) were equipped with a Ge Neutron Transmutation Doped (Ge-NTD) thermal sensor and operated as cryogenic detectors at about 10~mK. The achieved energy resolution was in the range of 100~keV FWHM at 1.3~MeV, and, while investigating the $^{210}$Pb content in the detector, no $^{210}$Po signal above the background level was observed. A limit on the $^{210}$Pb concentration in the detector was set $<$4~mBq/kg~\cite{4mBq} at 95\% C.L.

The difficulties in running such detectors are highlighted by the poor energy resolution. In fact operating a cryogenic detector in a superconducting state\footnote{Pb has a critical temperature of 7.2~K.} is not trivial when using a thermal sensor sensitive to thermal phonons, like a Ge-NTD. Furthermore the Debye temperature of the detector is about 80~K, which makes its heat capacity at low temperatures not favourable for low energy studies.

The performance of such detectors might be enhanced operating Pb with other classes of thermal sensors, suitable to read out superconducting absorber.

\section{Pure-Pb cryogenic detector}
The detector used for the measurement described in this work is one of the two samples already employed in~\cite{4mBq}. The absorber has a volume of 1~cm$^3$ and a mass of 10~g. This was extracted from a 23~kg Roman Pb ingot, without the need of any further processing.

The absorber was equipped with a Ge-NTD thermistor similar to the one used in~\cite{CUPID_detector}, but with reduced dimensions: 3.0$\times$1.0$\times$0.3~mm$^3$. The choice for a miniaturized thermal sensor is driven by the possibility to enhance the detector performance, namely the signal amplitude, while reducing the overall heat capacity of the system: absorber plus thermal sensor. The detector was also instrumented with a Si heater which was operated as Joule resistor, meant for the monitoring and the correction of temperature drifts of the absorber during the data taking. 

\begin{figure}[t]
\centering
\includegraphics[width=0.4\textwidth]{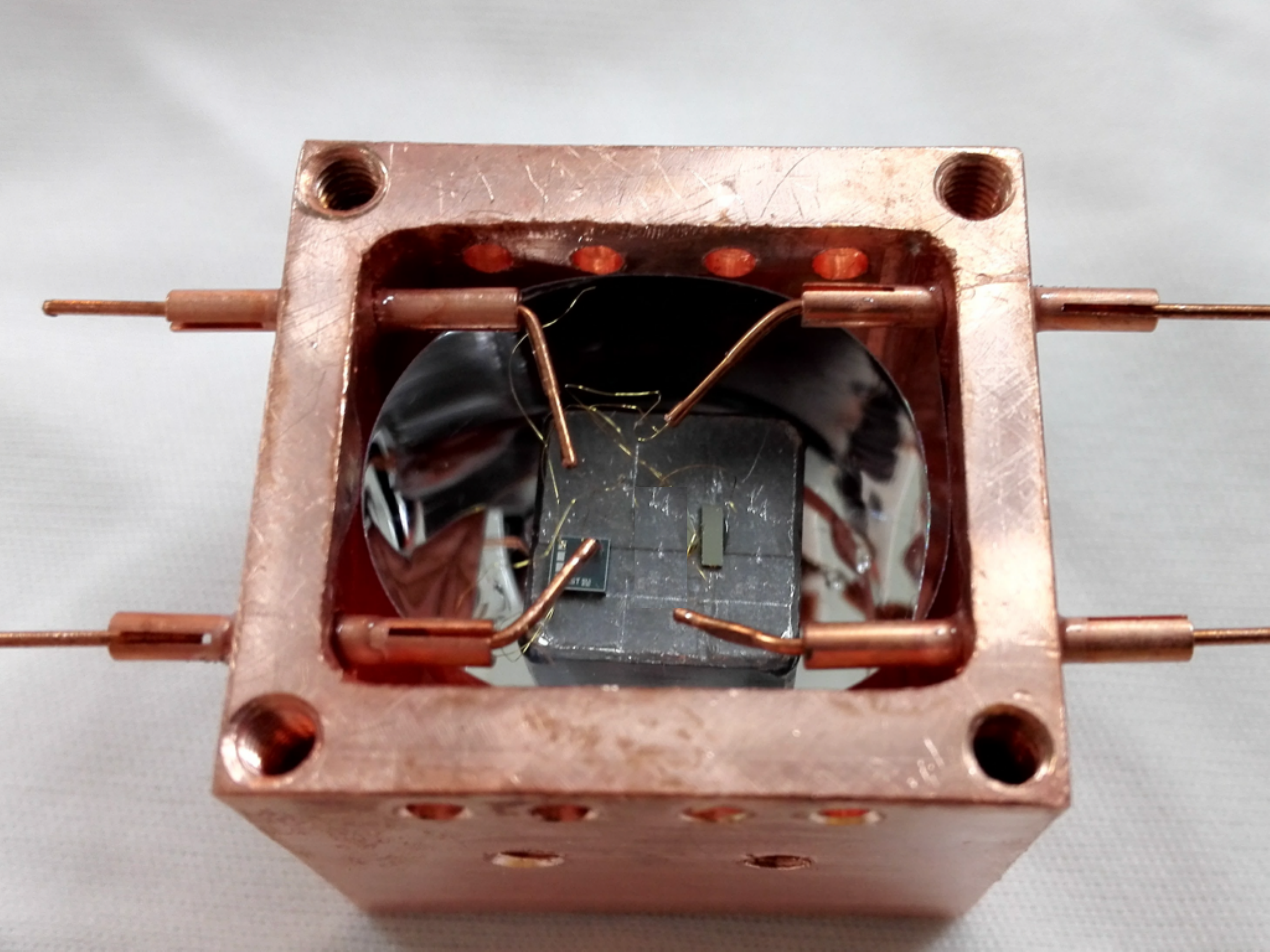}
\caption{Pure Pb cryogenic detector. The 1~cm$^3$ Pb absorber is visible inside the Cu housing. The Ge-NTD thermal sensor (on the right) and the Si Joule heater (on the left) are also visible.}
\label{fig:detector}
\end{figure}

The absorber was installed in a Cu housing and fixed to the support structure by means of epoxy resin. Copper was chosen not only because of its excellent thermal conductance at low temperatures, but also for minimizing possible surface radioactive backgrounds, given its excellent radiopurity. Our region of interest lies around 5.4~MeV (Q-value of $^{210}$Po), thus a possible background source can be ascribed to surface $\alpha$ decays as discussed in \cite{sticking}. A picture of the complete detector is shown in Fig.~\ref{fig:detector}.

The detector was installed in the low background CUPID R\&D cryogenic infrastructure located at the Laboratori Nazionali del Gran Sasso of INFN (Italy).

The thermal sensor was biased with a constant voltage through two 5~G$\mathrm{\Omega}$ load resistors in series, which ensured a constant current flow of 2.9~nA through the thermistor. Its working resistance was 2.7~M$\mathrm{\Omega}$, thus much lower than the load resistors in series. The acquired signals were amplified by custom-made low noise electronics~\cite{Elett} and then filtered by means of a 6-pole Bessel filter with cutting frequency of 500~Hz. Finally, the signals were fed into a NI-PXI-6284 18-bit ADC. The sampling frequency was 1~kHz and the waveforms were recorded when a software derivative trigger fired.

\section{Results}
The detector was operated for about 300~h, during which both background and calibration data were collected.
During calibration runs the detector was exposed to a $^{232}$Th source deployed next to the experimental set-up, but outside the cryogenic system. Characteristic monochromatic $\gamma$ emissions from the source were used for the calibration of the energy response of the detector, as well as for the monitoring of detector stability in time. Furthermore, a low intensity $^{232}$Th source was also installed nearby the detector, to study the detector response at lower energies.

The detector response to $\alpha$ interactions was not investigated in order not to spoil the detector background. In fact, this would have required the installation of a permanent $\alpha$ decay source directly facing the detector. This type of investigation was already performed with this very same detector in previous studies, in which the detector response was assumed to be the same for both $\alpha$ and $\beta/\gamma$ interactions.

\begin{figure}[h]
\centering
\includegraphics[width=0.5\textwidth]{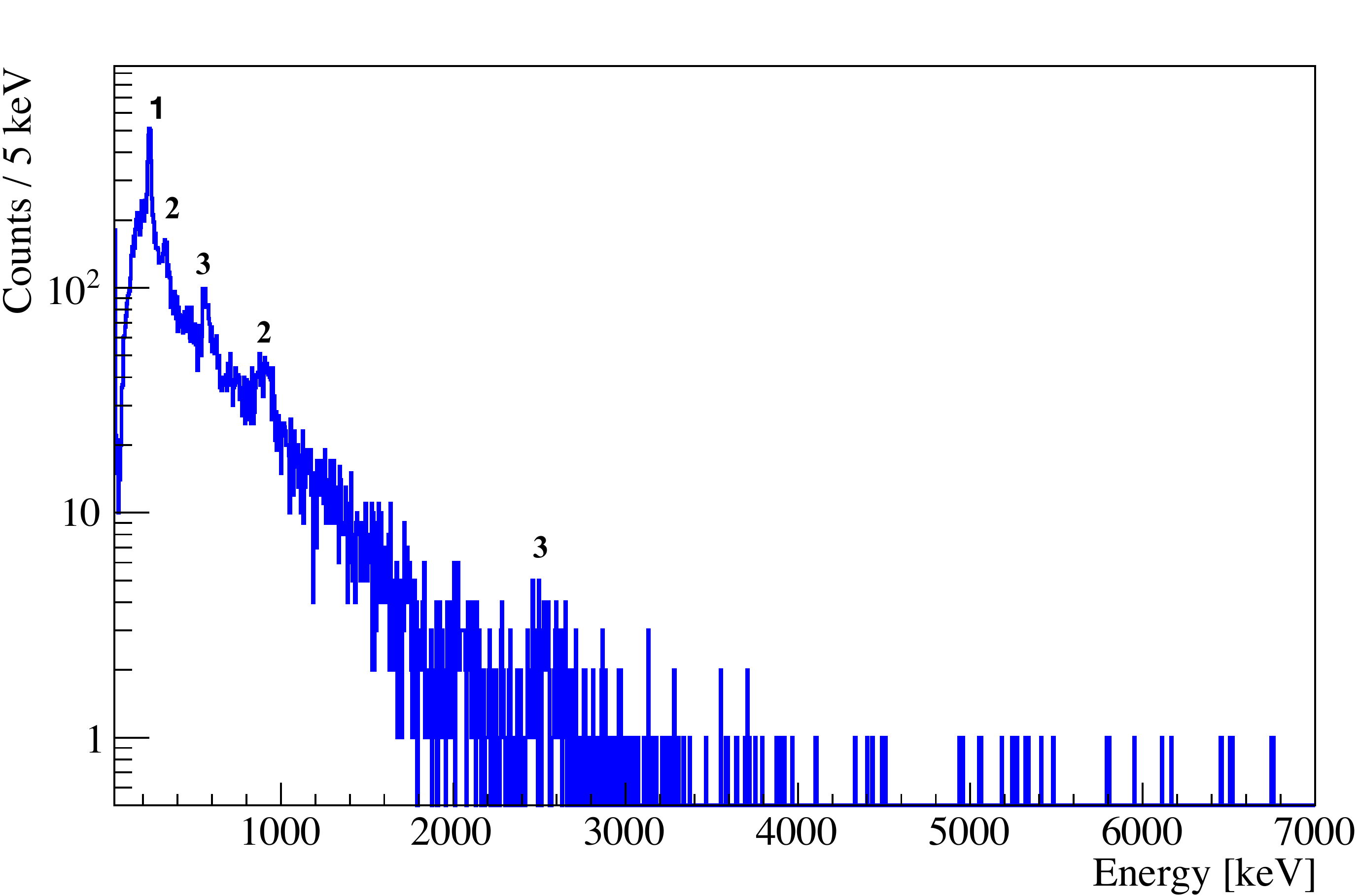}
\caption{Total background energy spectrum of a 1~cm$^3$ pure Pb cryogenic detector acquired over 229~h. The peaks are labeled as follows: (\textit{1}) $^{212}$Pb, (\textit{2}) $^{228}$Ac, and (\textit{3}) $^{208}$Tl.}
\label{fig:bkg}
\end{figure}

Background data were acquired for 229~h, when the detector was not exposed to calibration sources. The triggered data were studied by means of a two levels analysis system. Firstly, data were selected rejecting periods of noise excess due to interferences in the experimental set-up (i.e. electromagnetic disturbances or microphonic noise). Then, a second level data selection was performed, where by means of a pulse-shape analysis only particle events are selected. The total data selection efficiency is estimated to be 95$\pm$4\% at 2.6~MeV, the most intense high energy monochromatic $\gamma$-line in the acquired energy spectrum.

The results of the data selection analysis are shown in Fig.~\ref{fig:bkg}, reporting the total background energy spectrum.

\begin{figure}[h]
\centering
\includegraphics[width=0.5\textwidth]{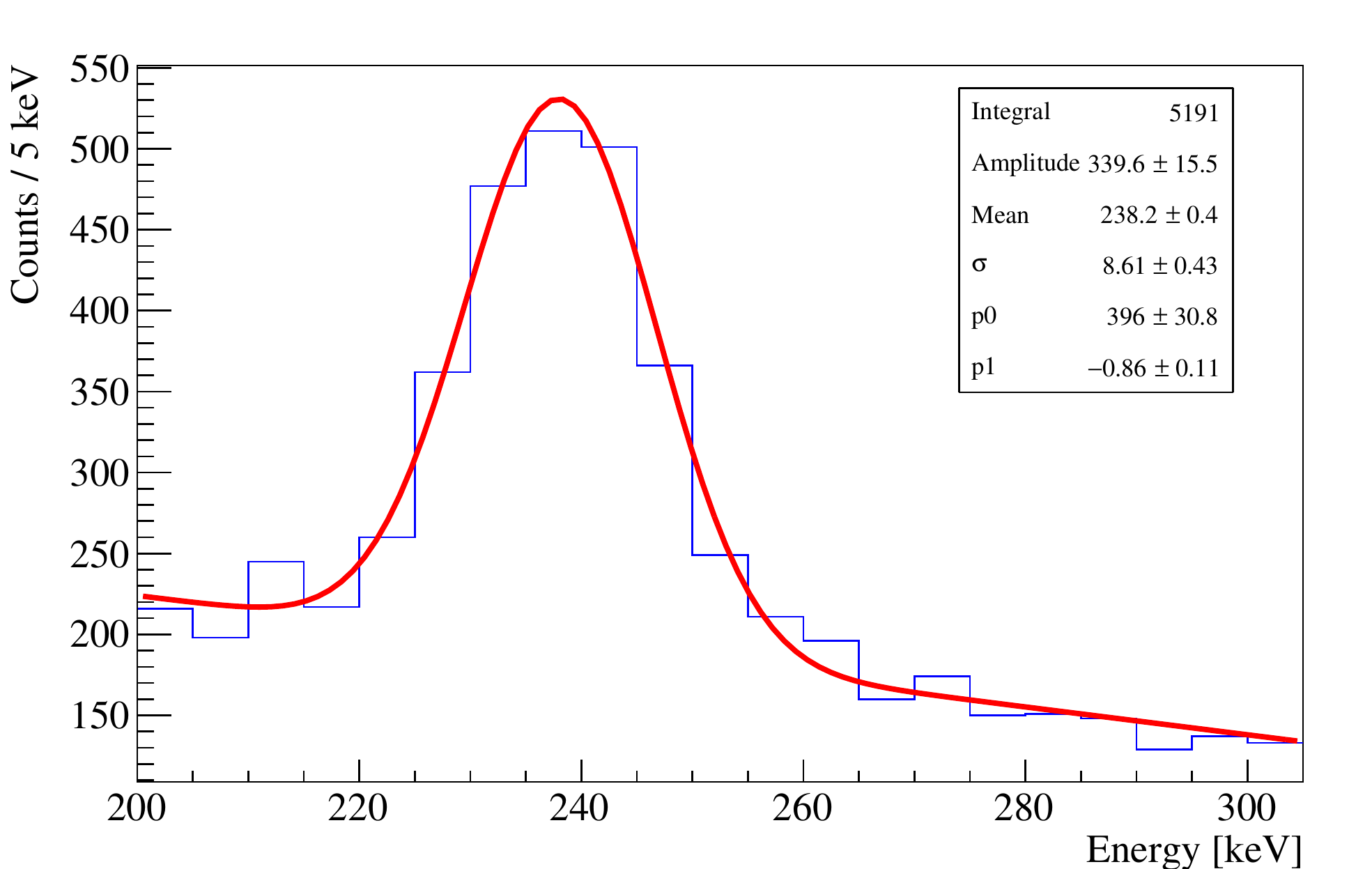}
\caption{Detector energy response at 238~keV. This signature is induced by a $\gamma$ interaction from $^{212}$Pb. The data are fitted using a convolution of a Gaussian function with a linear background.}
\label{fig:238}
\end{figure}

The energy spectrum features few monochromatic $\gamma$ lines, ascribed to the near $^{232}$Th calibration source. The FWHM resolution for the most prominent line, from $^{212}$Pb, at 238~keV, is 22$\pm$1~keV. At this energy scale the detector response is clearly Gaussian, as shown in Fig.~\ref{fig:238}.

\begin{figure}[h]
\centering
\includegraphics[width=0.5\textwidth]{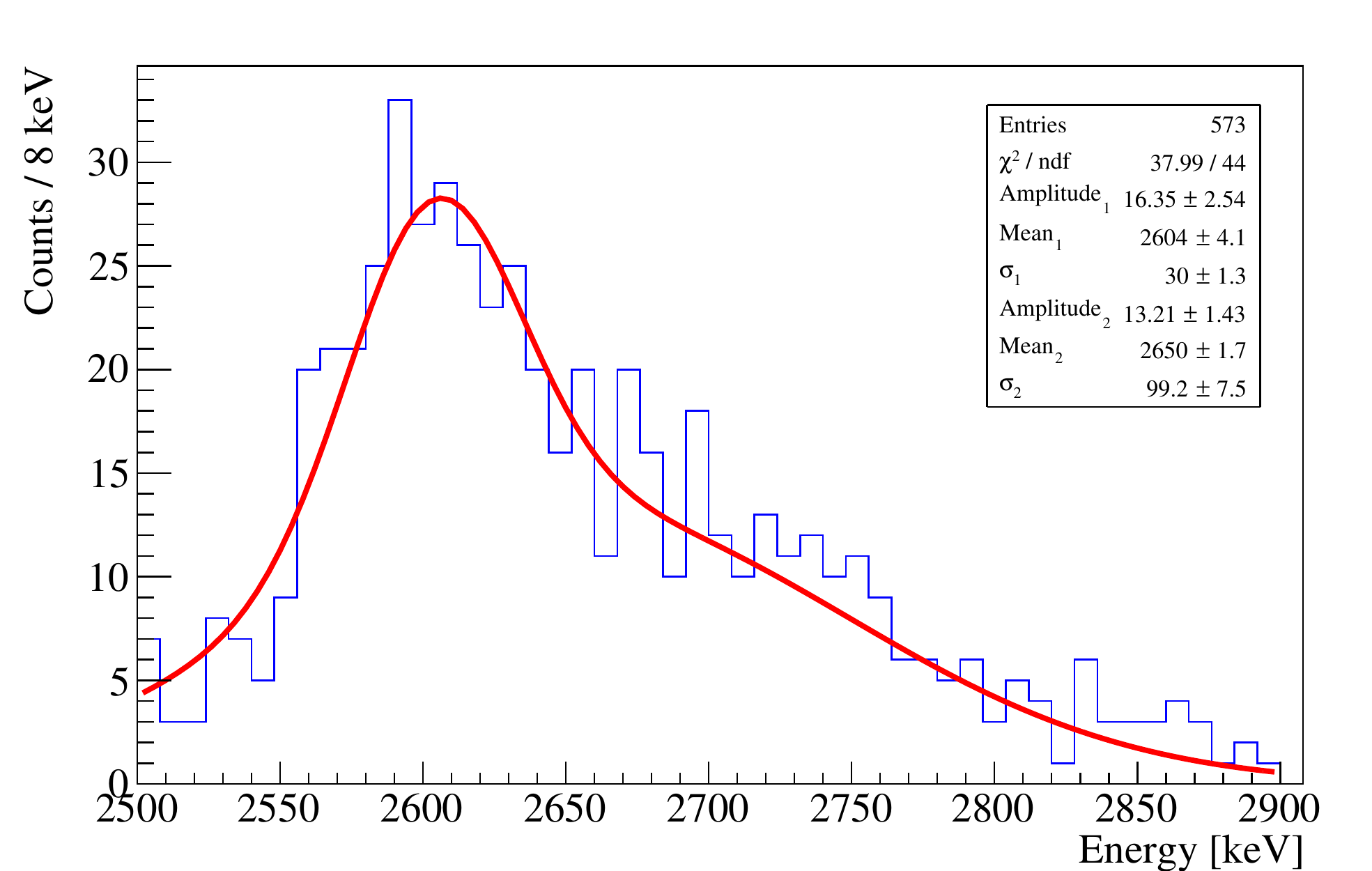}
\caption{Detector energy response at 2.6~MeV. This signature is induced by $\gamma$ interaction from $^{208}$Tl. The data are fitted using a double-Gaussian function.}
\label{fig:2615}
\end{figure}

For the detector energy calibration and the study of the detector response at higher energies, an intense $^{232}$Th source was deployed next to the cryogenic system. The detector response at the highest energy $\gamma$ line (2.6~MeV) is shown in Fig.~\ref{fig:2615}.
Different line shapes were studied, but the simplest that reproduced the detector response was a double-Gaussian. The reason for such a shape may rely on the non-uniformity of the detector response. In fact the detector is not a single crystal and possible position dependence effects could influence the response at high energies, due to multi-Compton interactions in the absorber. This behaviour was also observed in~\cite{Cuore}. The FWHM resolution at 2.6~MeV is 103.6$\pm$9.0~keV.

\begin{figure}[b]
\centering
\includegraphics[width=0.5\textwidth]{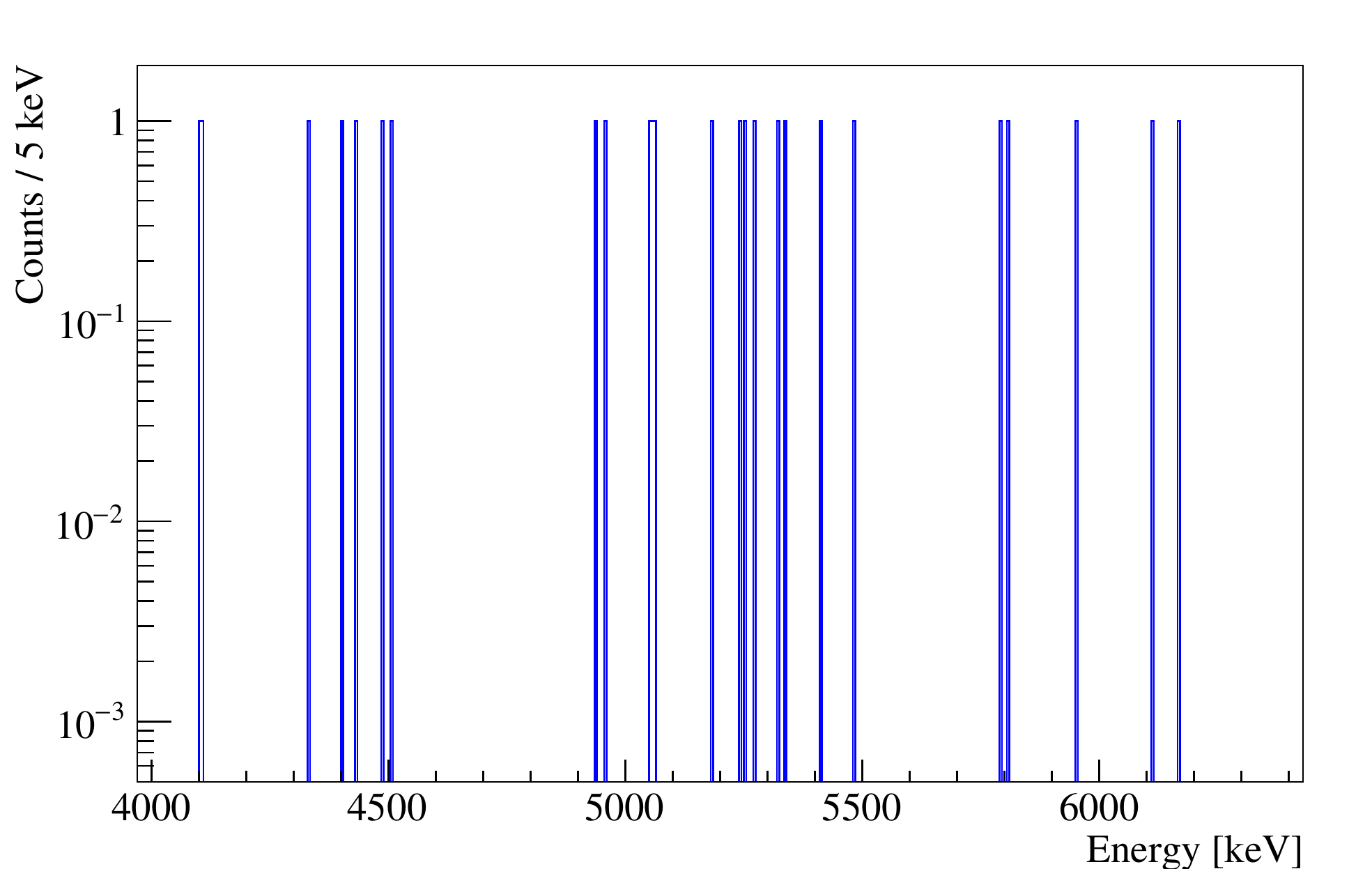}
\caption{Total background energy spectrum acquired over 229~h zoomed in the region of interest.}
\label{fig:zoom}
\end{figure}

As shown in Fig.~\ref{fig:bkg}, the detector did not feature any clear signature around 5.4~MeV, which is exactly the Q-value of $^{210}$Po. In first approximation we can assume that the detector energy resolution scales linearly with energy~\cite{Cupid}, for this reason we would expect the energy resolution at 5.4~MeV to be 200$\pm$17~keV.  In Fig.~\ref{fig:zoom}, a zoom in around the region of interest is shown.

We determine the $^{210}$Pb activity in our sample applying the Feldman-Cousin method~\cite{FC}. The signal is evaluated in the 200~keV RoI around the Q-value (4~events), while the background on the adjacent side bands (3~events). The upper limit at a 90\% C.L. is $<$715~$\mu$Bq/kg (5.6~events) at a 90\% confidence level. The result achieved with this set-up is about one order of magnitude better than any previous established limit. The detector counting rate in this region is extremely small, demonstrating the excellent radiopurity level of archeological Roman Pb. 

Using the same approach previously described, we were able to set upper limits on the concentration of $^{232}$Th and $^{238}$U decay chains, which are $<$0.44~mBq/kg and $<$0.34~mBq/kg, respectively. These values are not competitive with the ones reported in~\cite{Bkg_budget}, but it is a further confirmation of the purity of such valuable material.

\section{Conclusions}
In this work we operated a 1~cm$^3$ cryogenic detector made of pure archeological Roman Pb, one of the two absorbers that was used in~\cite{4mBq}. The low-background environment in which we ran the detector allowed us to study the internal radiopurity of the absorber. The most stringent limit on the concentration of $^{210}$Pb was established to be $<$715~$\mu$Bq/kg ($<$2.4$\times$10$^{-19}$~g/g). The absorber features an overall excellent radiopurity and no excess of $^{232}$Th and $^{238}$U decay chain products was observed; limits on their concentration were determined to be  $<$0.44~mBq/kg and $<$0.34~mBq/kg, respectively.

The excellent radiopurity level and the current performance as cryogenic detector make this material an interesting candidate for neutrino physics applications such as coherent neutrino-nucleus elastic scattering, but especially for the detection of supernova neutrinos, as discussed in~\cite{SN}. The performance as a cryogenic detector can be further optimized by the growth of a single-crystal absorber (e.g. melting and ri-crystallization of ancient Pb) and by the development of suitable thermal sensors (e.g. Ge-NTDs or Transition Edge Sensors).

\begin{acknowledgements}
Thanks are due to the LNGS mechanical workshop and in particular to E.~Tatananni, A.~Rotilio, A.~Corsi, and B.~Romualdi for continuous and constructive help in the overall set-up construction. Finally, we are especially grateful to M.~Perego and M.~Guetti for their invaluable help, and also to M. Nastasi for fruitful discussions.
\end{acknowledgements}

\end{document}